\documentclass[a4paper,11pt]{article}
\pdfoutput=1 % if your are submitting a pdflatex (i.e. if you have
\usepackage{epsfig}
\usepackage{graphicx}                  % provides \includegraphics command
\usepackage{ae}
\usepackage{amsmath}
\usepackage{amssymb}
\usepackage{graphics}
\usepackage{dcolumn}
\usepackage{bm}
\usepackage{lineno}
\usepackage{caption}
\usepackage{ragged2e}
\usepackage{subcaption}

\renewcommand{\thefootnote}{\fnsymbol{footnote}}
\title{\bf Long-term stability study of single-mask triple GEM detector: impact of continuous irradiation}
\date{}

\begin{document}

\maketitle
	\flushbottom
\vspace*{-1cm}
\centering

\author{\bf S.~Mandal,}
\author{\bf S.~Gope,}
\author{\bf S.~Das,}
\author{\bf S. Biswas$^*$}
\let\thefootnote\relax\footnotetext{$^*$Corresponding author. 

\hspace*{0.4cm}E-mail: saikat@jcbose.ac.in, saikat.ino@gmail.com }

\vspace*{0.5cm}

	{{Department of Physical Sciences, Bose Institute, EN-80, Sector V, Kolkata~-~700091, INDIA}

\vspace*{0.5cm}
\centering{\bf Abstract}
\justify

A study has been carried out to evaluate the performance stability of Gas Electron Multiplier (GEM) chamber prototypes in the laboratory using $^{55}$Fe radiation source with Argon and CO$_2$ gas mixture. This research focuses on the characterisation of the GEM detector's gain, efficiency (count rate with radioactive source), and energy resolution under varying operational conditions. A patch on the detector has been subjected to continuous and absolutely uninterrupted radiation for about 98 days. The gain and energy resolution of the detector are measured along with the ambient parameters temperature (t), pressure (p) and relative humidity (RH). In addition to that, the long-term behaviour of the count rate with a strong radioactive source  are also studied. This work is very relevant for Micro Pattern Gaseous Detectors (MPGD) such as GEM before installing on large experiment. The experimental setup, methodology, and results are presented in this article.



%\begin{keyword}
%% keywords here, in the form: keyword \sep keyword, up to a maximum of 6 keywords
%keyword 1 \sep keyword 2 \sep keyword 3 \sep keyword 4



%\end{keyword}
\vspace*{0.25cm}


%\tableofcontents



\section{Introduction}
Gas Electron Multiplier (GEM) technology introduced by Fabio Sauli in 1997~\cite{Sauli}, as an effective detection system has become a major breakthrough in High Energy Physics (HEP). GEM, one of the members of the Micro-Pattern Gas Detector (MPGD) family, have been noted to be preferred due to its ability to handle high rate ($\sim$1~MHz/mm$^2$) and good position resolution ($\sim$100~$\mu$m)~\cite{Sauli,Buzulutskov,Ketzer}. This combination of properties makes them fundamentally better suited to the systems located in high-radiation areas in any HEP experiment. Within the framework of the future Compressed Baryonic Matter (CBM) experiment at the Facility for Antiproton and Ion Research (FAIR), Darmstadt, Germany, GEM detectors are chosen in the first and second stations of the Muon Chamber (MuCh), where their features of muon tracking with high efficiency will be explored~\cite{Galatyuk,cbm,fair,rama}. In the third and fourth stations of CBM-MuCh, Straw Tube detectors or single gap Resistive Plate Chamber (RPC) system are proposed options as in these stations comparatively lower particle rate is expected ($\sim$13-15~kHz/cm$^2$ and $\sim$7.5-10~kHz/cm$^2$ respectively)~\cite{sharma2023}.  In recent years, significant studies are being carried out concerning the long-term stability of the GEM detectors that have been exposed to continuous radiation. The stability in the  gain, energy resolution, efficiency is essential for long-term use of a gaseous detector in harsh radiation environment.

This study deals with the performance stability of the GEM chamber prototypes with gas mixture of Argon and CO$_2$ in a 70/30 volume ratio and extended radiation exposures. The results of an investigation conducted over 3 months without any interruption, during which, a GEM chamber is subject to continuous irradiation with 5.9~keV $^{55}$Fe X-rays at around 220~kHz is reported. The examination shows several developments in the chamber's performance, particularly concerning the stabilisation of the bias current, in addition to the connection with the fluctuation of the gain and energy resolution. It serves as a bigger picture for the development of the device and the process of GEM detectors in future HEP experiments. In this article, section 2 describes the configuration of the detector and the experimental setup. Detailed specifications of the detector and the operational framework are explained in section 3. Section 4 covers the detail results. The study is summarised in section 5.

\section{Detector configuration and experimental setup}\label{det_conf} 
%%%%%%%%%%%%%%%%%%%%%%%%%%%%%%%%%%%%%%%%%%%%%%%%%%%
\begin{figure}[htb!]
	\begin{center}
		\includegraphics[scale=0.4]{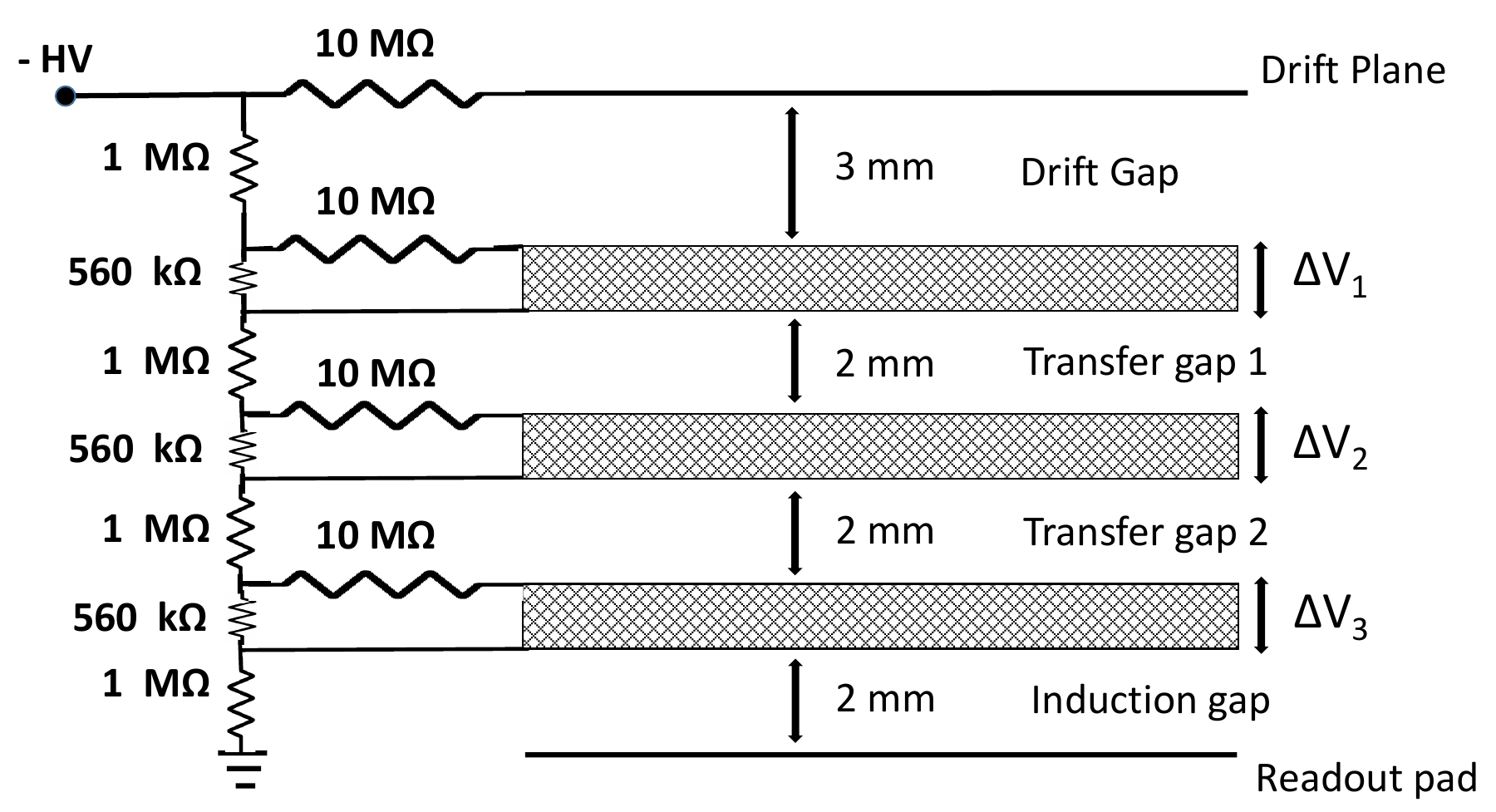}
		\caption{\label{fig1} Schematic of the high voltage (HV) distribution resistive chain network and different planes of the SM triple-GEM detector~\cite{s_chatterjee_charging_up_2}. }\label{fig1}
	\end{center}
\end{figure}
%%%%%%%%%%%%%%%%%%%%%%%%%%%%%%%%%%%%%%%%%%%%%%%%%%%

%%%%%%%%%%%%%%%%%%%%%%%%%%%%%%%%%%%%%%%%%%%%%%%%%%%
\begin{figure}[htb!]
	\begin{center} 
	\includegraphics[scale=0.4]{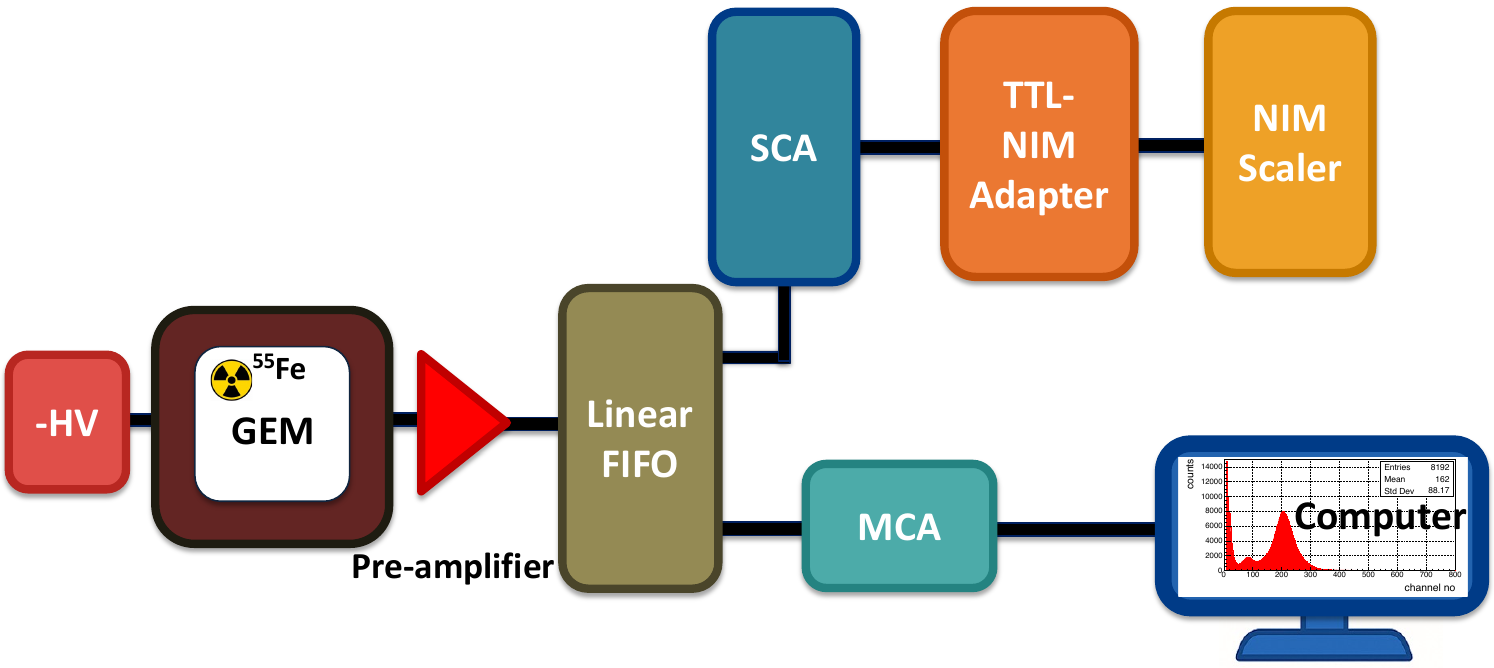}
	\caption{ Schematic representation of the electronic circuit \cite{mandal_stability}.}\label{fig2}
	\end{center}
\end{figure}
%%%%%%%%%%%%%%%%%%%%%%%%%%%%%%%%%%%%%%%%%%%%%%%%%%%

%%%%%%%%%%%%%%%%%%%%%%%%%%%%%%%%%%%%%%%%%%%%%%%%%%%
\begin{figure}[htb!]
	\begin{center} 
	\includegraphics[scale=0.5]{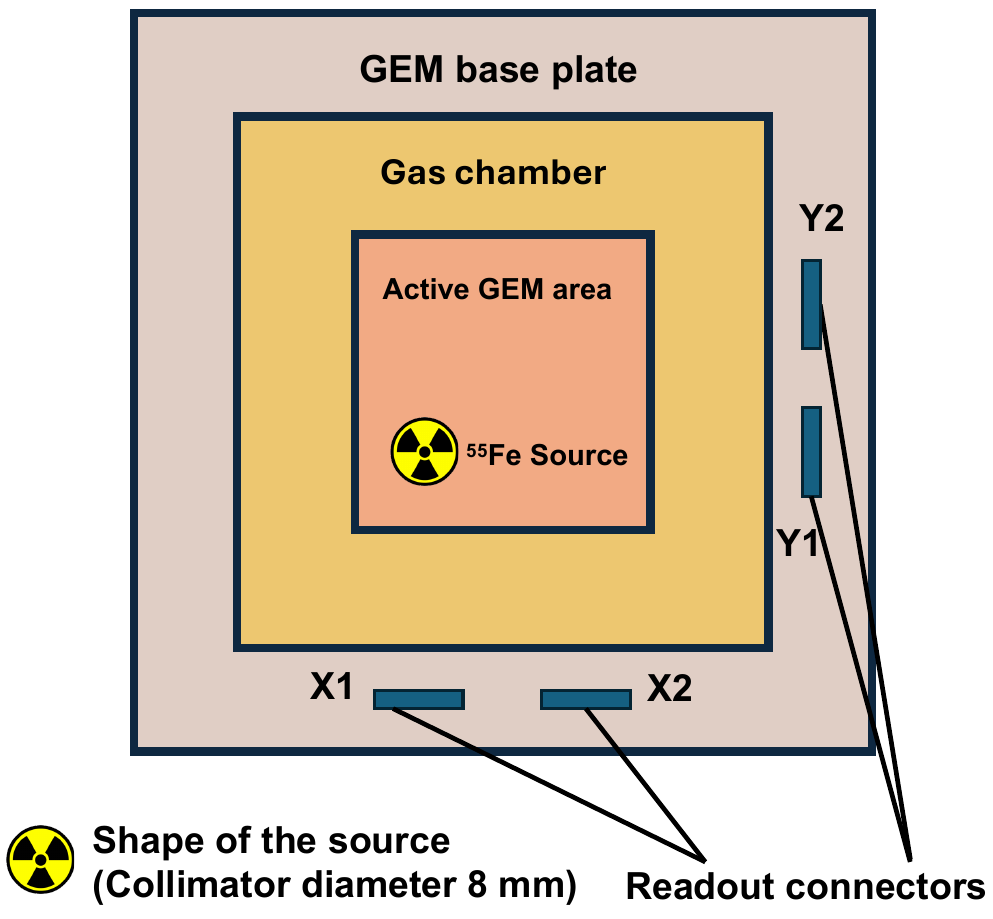}
	\caption{Schematic of the position of the radioactive source on the chamber.}\label{readout}
	\end{center}
\end{figure}
%%%%%%%%%%%%%%%%%%%%%%%%%%%%%%%%%%%%%%%%%%%%%%%%%%%

Standard GEM foil, consists of 50~$\mu$m thick Kapton film with 5~$\mu$m Copper cladding on both sides. A large number of holes ($\sim$5$\times$10$^4$ holes/cm$^2$) are pierced in the film using the photo-lithographic technique. Depending on the technique used, the GEM foils are classified in two categories, Double Mask (DM) or Single Mask (SM) GEM\cite{Sauli,biswas2015,oliveira2013,duarte2009,karadzhinova2015,das2016,shah2019,bachmann1999,s_chatterjee_charging_up_2}. The SM triple-GEM detector prototype used in this work is having dimension of 10~$\times$~10~cm$^2$ and in a 3-2-2-2 gap configuration in millimeter (respectively the drift gap, transfer gap 1, transfer gap 2 and induction gap). The GEM foils and other detector components are obtained from CERN. Figure~\ref{fig1} shows the configuration of the different planes and the resistive chain network to bias the GEM layers and drift plane. 10~M$\Omega$ protection resistors are placed in the drift plane and top layers of every GEM foil. While 560~k$\Omega$ parallel divider resistors are used across each GEM foil, as shown in Figure~\ref{fig1}, 1~M$\Omega$ parallel divider resistors are connected across the the drift, transfer, and induction gaps. A unique 10~$\times$~10~cm$^2$ plane made with 512 copper strips (256 X and 256 Y segments) is used as the readout plane. In this work, 128 strips from each X and Y side are combined using a sum-up board provided by CERN and signals are collected by a single Lemo output. For this chamber, four of these sum-up boards are used - 2 for X-side and 2 for Y-side.

In the electronic circuit, a charge-sensitive preamplifier (VV50-2), with a gain of 2~mV/fC and a shaping time of 300~ns~\cite{preamplifier}, is used to receive the signals from the sum-up board. After that, a linear Fan-In Fan-Out (Linear-FIFO) module processes this analog signal. The Linear-FIFO can split a single analog signal into four signals of exactly same amplitude. One of the outputs from the Linear-FIFO is fed to a Multi-Channel Analyser (MCA) to store the X-ray spectra on a desktop computer. Another output from the FIFO is fed into a Single Channel Analyser (SCA), which is used in the integral mode to serve as a discriminator and the signals above the threshold is counted using a NIM scaler. During the measurements, the SCA threshold is kept at 0.4~V. The schematic view of the electronic circuit diagram is shown in figure~\ref{fig2}. The details of the detector specification, configuration of the resistor chain and experimental details are provided in Table~\ref{table1}\cite{mandal_2024}.

In this study, a gas mixture of Argon (Ar) and CO$_2$ in the 70/30 volume ratio is used. A constant flow rate of $\sim$~3.5-4l/hr is maintained using a V{\"o}gtlin gas flow meter. A $^{55}$Fe X-ray source having characteristic energy of 5.9~keV is used to radiate the chamber continuously. To irradiate a specific region on the detector by X-rays, a circular collimator of diameter 8~mm is used to ensure the exposure area on the detector $\sim$~50~mm$^2$ for all the measurements\cite{Roy_thesis,Sen_thesis,Chatterjee_thesis,adak,s_roy_gain_calculation,uniformity_1,chatterjee_2020,s_chatterjee_charging_up_1,chatterjee_2023_rh,chatterjee_2023_charge}. The position of the radioactive source on the detector is shown in figure~\ref{readout}.

%%%%%%%%%%%%%%%%%%%%%%%%%%%%%%%%%%%%%%%%%%%%%%%%%%%%%%%%%%%%%%%%%%%
\begin{table}[htb!]
	\centering
	\small
	\resizebox{0.83\linewidth}{!}{%
		\begin{tabular}{|l|l|}
			\hline
			\textbf{Item} & \textbf{Specification} \\
			\hline
			\multicolumn{2}{|c|}{\textbf{Detector description}} \\
			\hline
			Detector dimension & 10 $\times$ 10 cm$^2$ \\
			GEM foils & Made with single-mask technique \\
			No. of foils & 3 \\
			\hline
			\multicolumn{2}{|c|}{\textbf{Different gaps}} \\
			\hline
			Drift gap & 3 mm \\
			Transfer gap 1 & 2 mm \\
			Transfer gap 2 & 2 mm \\
			Induction gap & 2 mm \\
			\hline
			\multicolumn{2}{|c|}{\textbf{Value of resistors in divider chain}} \\
			\hline
			Protection resistors & 10 M$\Omega$ \\
			Parallel divider resistances in drift gap, transfer gaps and induction gap & 1 M$\Omega$ \\
			Parallel divider resistances across each GEM foil & 560 k$\Omega$ \\
			\hline
			\multicolumn{2}{|c|}{\textbf{Electrical parameters at typical HV of - 4700 V}} \\
			\hline
			Typical bias current & $\sim$720 $\mu$A \\
			$\Delta V$ across GEM foil & $\sim$403.2 V \\
			Drift field & $\sim$2.6 kV/cm \\
			Transfer fields & $\sim$3.6 kV/cm \\
			Induction field & $\sim$3.6 kV/cm \\
			\hline
			\multicolumn{2}{|c|}{\textbf{Readout}} \\
			\hline
			No. of strips & 512 strips (256 X, 256 Y) \\
			No. of sum-up boards & 4 \\
			Each sum-up board can add & 128 strips \\
			\hline
			\multicolumn{2}{|c|}{\textbf{Specification of the charge sensitive preamplifier}} \\
			\hline
			Gain & 2 mV/fC \\
			Shaping time & 300 ns \\
			\hline
			\multicolumn{2}{|c|}{\textbf{Gas}} \\
			\hline
			Gas mixture & Ar/CO$_2$: 70/30 (volume ratio) \\
			Flow rate & 3.8 l/hr \\
			\hline
			\multicolumn{2}{|c|}{\textbf{Radiation Source}} \\
			\hline
			Name of the X-ray source & $^{55}$Fe \\
			Activity & 6.25 mCi \\
			Energy of the X-ray & 5.9 keV \\
			Rate of the X-ray & $\sim$220 kHz \\
			\hline
			\multicolumn{2}{|c|}{\textbf{Exposed area and position}} \\
			\hline
			Exposed area & 50 mm$^2$ \\
			Exposed position & 42 (imaginary) \\
			\hline
		\end{tabular}
	}
	\caption{\label{table1}Experimental details of the GEM detector and associated setup.}
\end{table}

%%%%%%%%%%%%%%%%%%%%%%%%%%%%%%%%%%%%%%%%%%%%%%%%%%%%%%%%%%%%%%%%%%%

\section{Description of the observables} \label{math}

%%%%%%%%%%%%%%%%%%%%%%%%%%%%%%%%%%%%%%%%%%%%%%%%%%%
\begin{figure}[htb!]
	\begin{center}
		\includegraphics[scale=0.6]{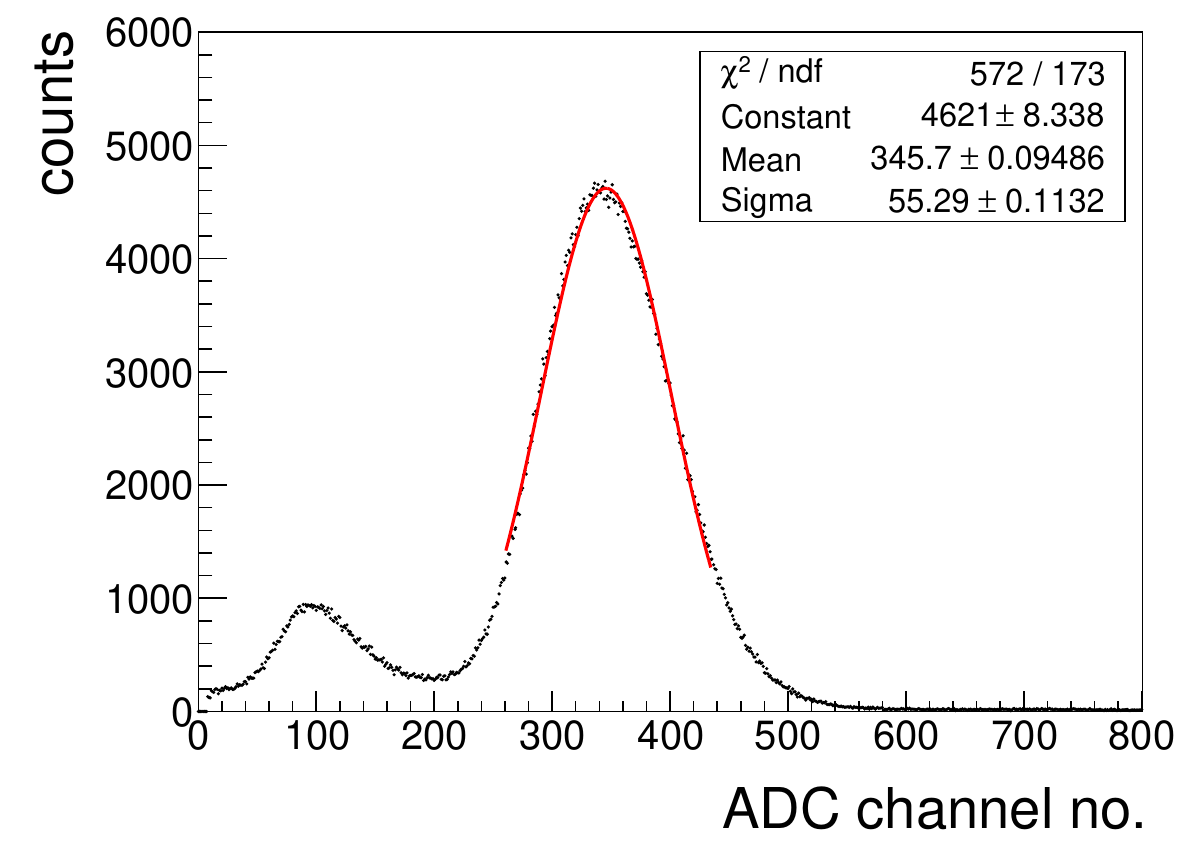}
		\caption{(Colour online) Typical $^{55}$Fe spectrum in Ar/CO$_2$ (70/30 volume ratio) gas mixture at -~4700~V. The $\Delta V$ across each of the GEM foils is $\sim$~403.2~V.}\label{fig2}
	\end{center}
\end{figure}
%%%%%%%%%%%%%%%%%%%%%%%%%%%%%%%%%%%%%%%%%%%%%%%%%%%

Irradiation of the chamber is done using a $^{55}$Fe source of X-rays as explained in Section~\ref{det_conf}. The typical spectrum obtained at a HV of -~4700~V corresponding to a $\Delta{V}$~=~403.2~V on each GEM foil is displayed in figure~\ref{fig2}. The spectrum shows a 2.9 keV escape peak, and a clear main peak at 5.9 keV. The 5.9~keV peak is fitted with a Gaussian function to calculated the gain and energy resolution of the chamber. The total output charge for further analysis is estimated by the mean value of this Gaussian fit, preamplifier gain, and MCA calibration factor.

The equations regarding gain and energy resolution and charge accumulation with their derivation are also included together with the specifications of GEMs and the experimental arrangements made in this work \cite{mandal_2024}. These detailed specifications are useful in ascertaining the operating conditions of the detected system. 

According to the standard definition, the gain of the detector can be expressed as

\begin{equation}
\begin{aligned}
gain  
& = \frac{output \; charge}{input \: charge} 
= \frac{\frac{Pulse \; hight}{2 \: mV} fC}{no.\; of \; primary \; electrons\; \times \; eC} 
\end{aligned}
\end{equation}

The pulse height is obtained using the calibration factor of the MCA. The MCA is calibrated with a known input signal from a pulse generator. The relation between the pulse height and the corresponding mean MCA channel number is given by

\begin{equation}
\begin{aligned}
Pulse \; height \; (V) = MCA \; Channel \; no. \times 0.0014 \; + \;  0.1428
\end{aligned}
\end{equation}

Further details of the MCA calibration procedure can be found in Ref.~\cite{Roy_thesis}.

The energy resolution of the detector is determined from the full width at half maximum (FWHM) of the Gaussian-fitted energy spectrum and is calculated using

\begin{equation}
\centering
Energy \; resolution = \frac{sigma \times 2.355}{mean}
\end{equation}

where $\sigma$ and the mean are obtained from the Gaussian fit to the energy spectrum.

For stability studies of a GEM detector, the energy spectra along with ambient parameters such as temperature ($t$ in $^\circ$C), pressure ($p$ in mbar), and relative humidity ($RH$ in \%) are recorded at regular intervals. The gain ($G$) of a gaseous detector depends strongly on the ratio of absolute temperature ($T = t + 273$) to pressure ($p$), and follows the relation

\begin{equation}
\centering
G (T/p) = A e^{(B \frac{T}{p})}\label{correlation}
\end{equation}

where $A$ and $B$ are constants obtained from fitting.

To examine gain stability, the gain as a function of $T/p$ is fitted using Eq.~\ref{correlation}, and the parameters $A$ and $B$ are extracted. The measured gain is then normalized using

\begin{equation}
\centering
normalised \; gain = \frac{measured \; gain}{A e^{(B \frac{T}{p})}}\label{norm_g}
\end{equation}

The normalized gain is plotted as a function of the total accumulated charge per unit irradiated area of the detector, which is proportional to time. The accumulated charge per unit area ($dq/dA$) at a given time is calculated as

\begin{equation}
\centering
\frac{dq}{dA} = \frac{r \times n \times e \times G \times dt}{dA}\label{chperarea}
\end{equation}

where $r$ is the measured rate over a specific detector area, $dt$ is the time interval in second, $n$ is the number of primary electrons produced per incident X-ray, $e$ is the electronic charge, $G$ is the gain, and $dA$ is the irradiated area.

Similarly, the dependence of energy resolution on $T/p$ is described using an exponential function

\begin{equation}
\centering
energy \; resolution = A' e^{(B' \frac{T}{p})}\label{correlation_er}
\end{equation}

where $A'$ and $B'$ are fitting constants.

The measured energy resolution is then normalized using

\begin{equation}
\centering
normalised \; energy \; resolution =
\frac{measured \; energy \; resolution}{A' e^{(B' \frac{T}{p})}}\label{norm_er}
\end{equation}
%%%%%%%%%%%%%%%%%%%%%%%%%%%%%%%%%%%%%%%%%%%%%%%%%%%

%\begin{figure}
%\centering 
%\includegraphics[width=0.4\textwidth]{electronic.png}		
%\caption{Schematic of the electronic circuit used for data acquisition.}
%\label{fig1}%
%\end{figure}

%%\subsection{Subsection title}

%\begin{figure}
%	\centering 
%	\includegraphics[width=0.4\textwidth, angle=-90]{Results_in_Physics.pdf}	
%%	\label{fig_mom0}%
%\end{figure}

%A random equation, the Toomre stability criterion:

%\begin{equation}
%    Q = \frac{\sigma_v \times \kappa}{\pi \times G \times \Sigma}
%\end{equation}

\section{Results}\label{res} 
Results of long-term stability test of the SM triple-GEM detector is described here in detail. Initially, the HV is ramped up slowly and set at -~4500~V. X-ray of a rate of $\sim$~220~kHz is made to expose to an area of 50~mm$^2$ (corresponding to a flux of 0.4~MHz/cm$^2$) and the spectra are recorded for every 1 minute. The temperature (t) [absolute temperature T = t+273], pressure (p) and relative humidity (RH) are also recorded for every 1 minute interval using a data logger made in-house \cite{sahu,mandal_datalogger}. The value of the applied voltage and divider current are recorded manually from the HV power supply itself. The applied voltage and the divider current as a function of time which is one of the key observables in this study is shown in Figure~\ref{fig3}, the current as a function of time for the first 1 hour after applying HV is shown in the inset. The other observables such as count rate, energy resolution, gain, gas flow rate, relative humidity, pressure, temperature and T/p along with measured applied voltage and divider current as a function of time are shown at a glance in Figure~\ref{fig4}. 

%%%%%%%%%%%%%%%%%%%%%%%%%%%%%%%%%%%%%%%%%%%%%%%%%%%
\begin{figure}[htb!]
	\begin{center} 
		\includegraphics[scale=0.6]{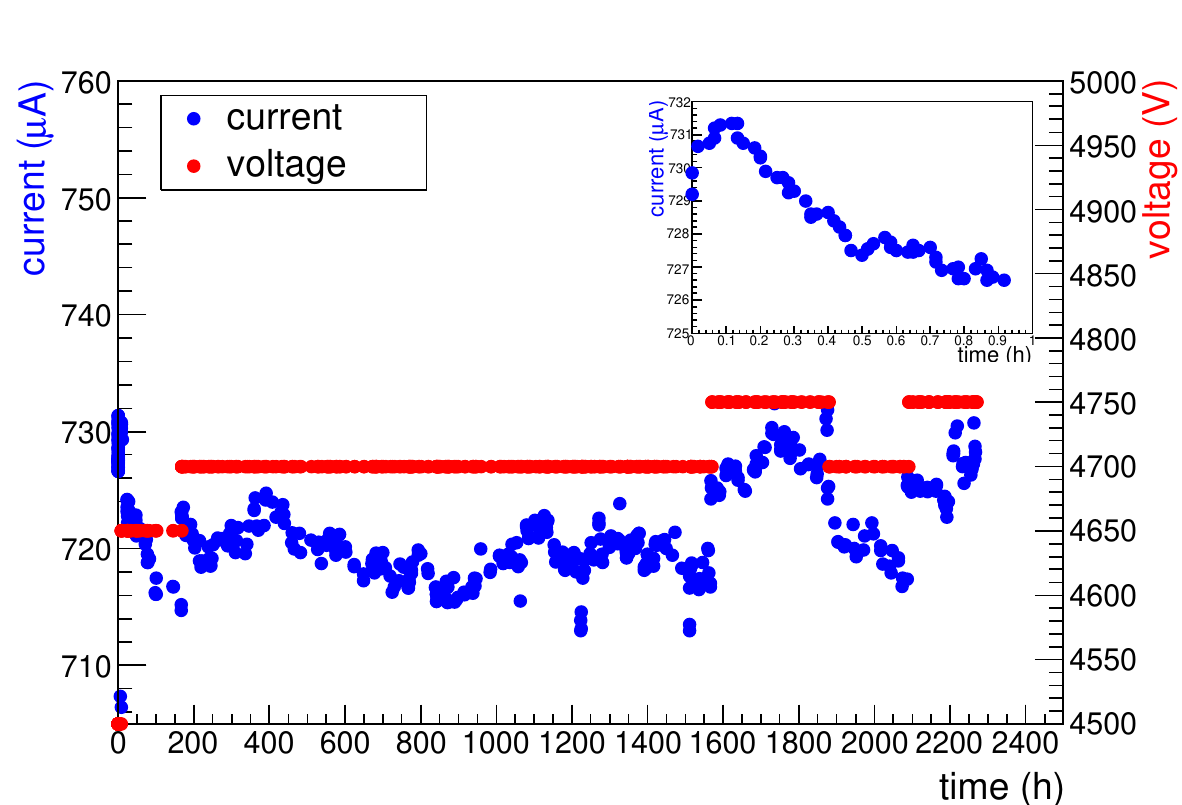}
		\caption{(Colour online) Applied voltage and divider current as a function of time. The current as a function of time for the first 1 hour after applying HV is shown in the inset.}\label{fig3}
	\end{center}
\end{figure}
%%%%%%%%%%%%%%%%%%%%%%%%%%%%%%%%%%%%%%%%%%%%%%%%%%%

%%%%%%%%%%%%%%%%%%%%%%%%%%%%%%%%%%%%%%%%%%%%%%%%%%%
\begin{figure*}[htbp]
	\begin{center} 
		\includegraphics[scale=0.630]{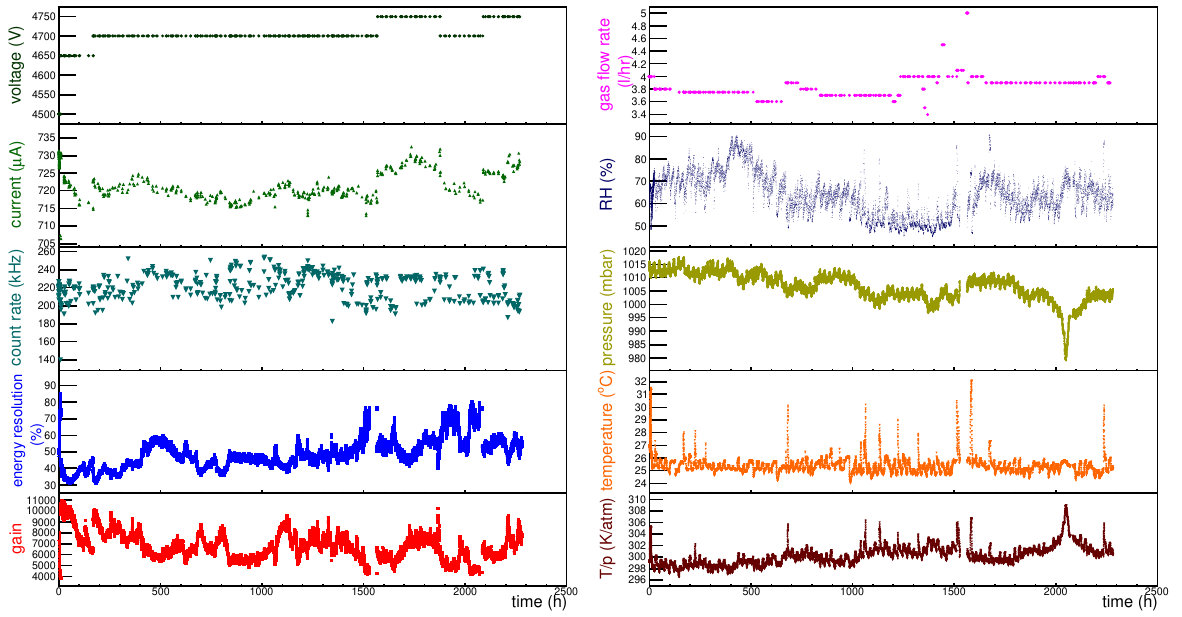}
		\caption{(Colour online) The applied voltage, measured divider current, count rate, energy resolution, gain, gas flow rate, relative humidity, pressure, temperature and T/p as a function of the time.}\label{fig4}
	\end{center}
\end{figure*}
%%%%%%%%%%%%%%%%%%%%%%%%%%%%%%%%%%%%%%%%%%%%%%%%%%%

It is observed from Figure~\ref{fig3} that immediately after applying HV the current increases rapidly, reaches a maximum and then starts to decrease. The shape of $^{55}$Fe X-ray spectra recorded after different time from the application of HV are shown in Figure~\ref{fig_sel_spec}.

%%%%%%%%%%%%%%%%%%%%%%%%%%%%%%%%%%%%%%%%%%%%%%%%%%%
\begin{figure}[htb!]
	\begin{center}
		\includegraphics[scale=0.65]{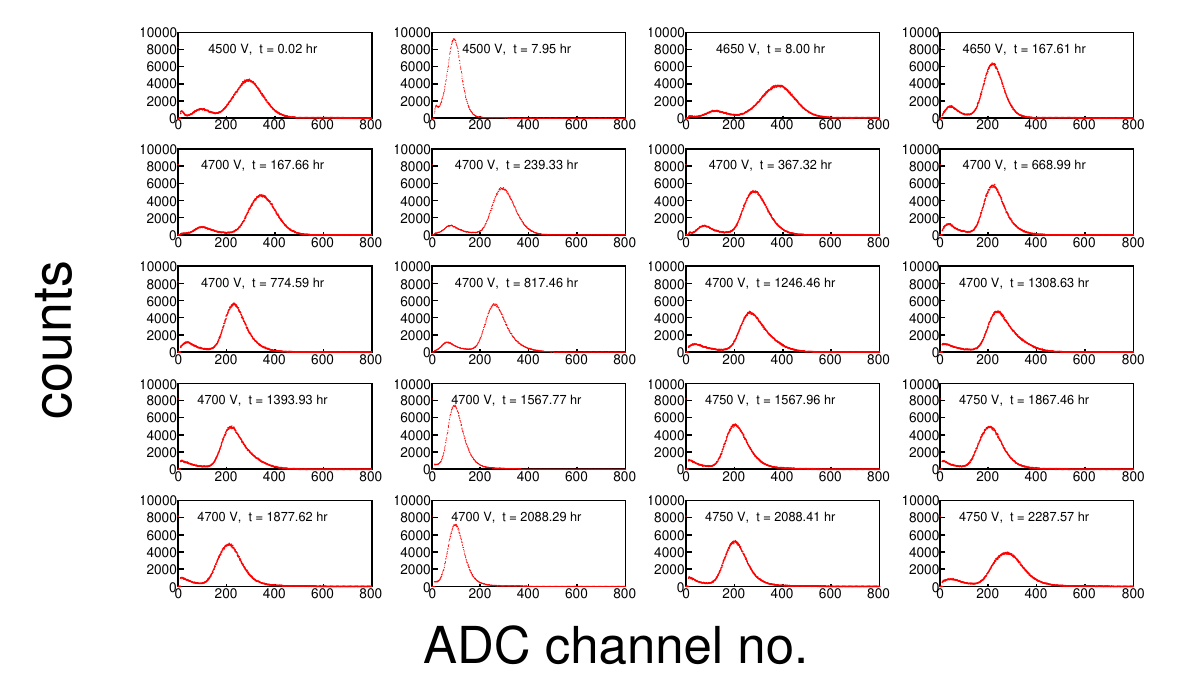}
		\caption{(Colour online) $^{55}$Fe spectra in different selected time.}\label{fig_sel_spec}
	\end{center}
\end{figure}
%%%%%%%%%%%%%%%%%%%%%%%%%%%%%%%%%%%%%%%%%%%%%%%%%%%

It is observed from Figure~\ref{fig4} that initial charging up effect could not be seen because the conditioning of the detector takes time, much larger than the charging up time \cite{mandal_2024}. From Figure~\ref{fig4} it is seen that the gain is gradually decreasing over time and the energy resolution is changing in such a way which is anti-correlated with the gain. Bias current also varies with time which may affect the gain.  
%It is clearly seen that when the current decreases, the gain also decreases. 
% However, variation is observed in voltage, temperature, pressure, gas flow rate, count rate and RH.

%%%%%%%%%%%%%%%%%%%%%%%%%%%%%%%%%%%%%%%%%%%%%%%%%%%
\begin{figure}[htb!]
	\begin{center} 
		\includegraphics[scale=0.630]{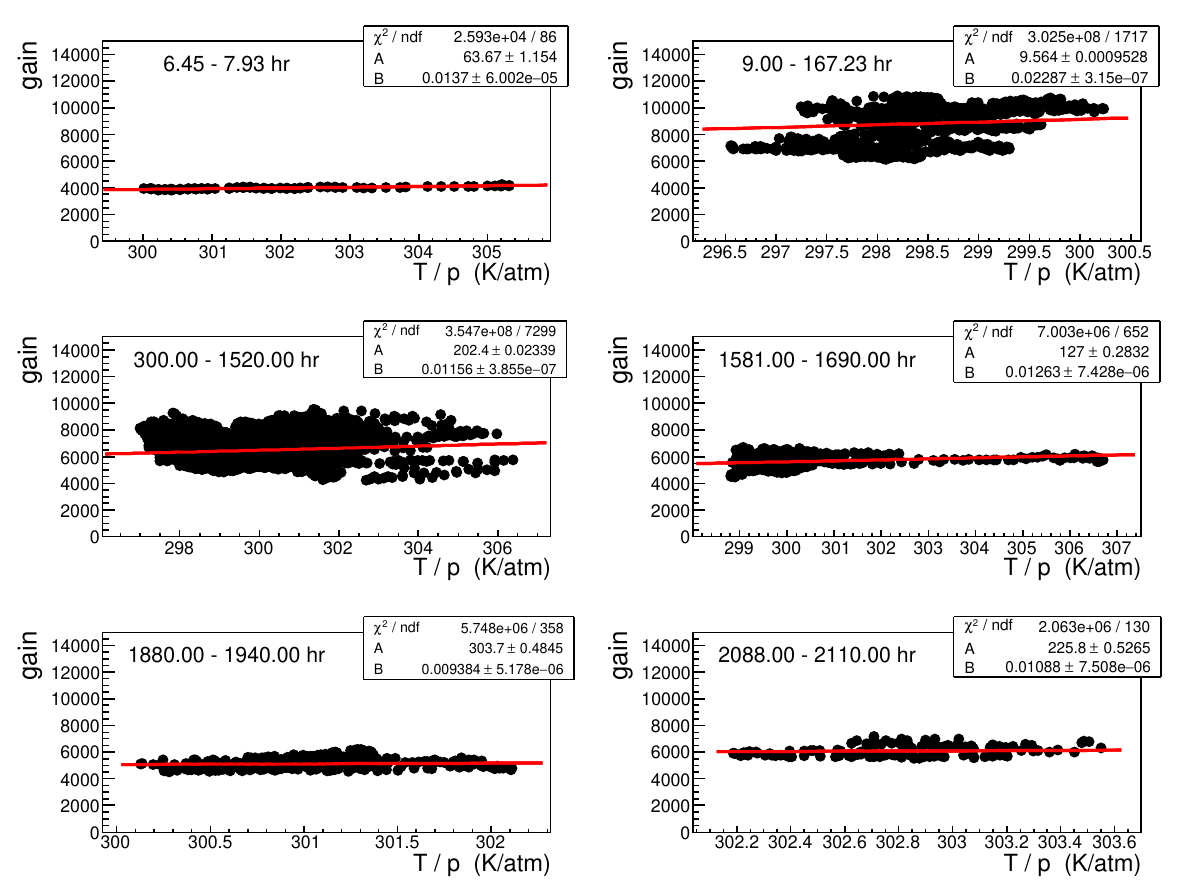}
		\caption{(Colour online) Gain and T/p correlation plot for different time zones.}\label{fig5}
	\end{center}
\end{figure}
%%%%%%%%%%%%%%%%%%%%%%%%%%%%%%%%%%%%%%%%%%%%%%%%%%%
%\vspace{-0.6cm}

The current at the starting point (at HV of -~4500~V) is measured to be about 729~$\mu$A which increases quickly to a value of $\sim$~731.5~$\mu$A and this is the maximum for this run after which it starts to decrease. After about 8 hours of operation, the current decreased to about 706~$\mu$A and the applied HV is increased to -~4650~V. At that point, the divider current is again raised to a value $\sim$~729~$\mu$A. Subsequently, as time progresses, the current drops to $\sim$~715~$\mu$A but slowly, in $\sim$~167~hours. Since the current fluctuates with time and shows an overall decrease, the gain also fluctuates and gradually decreases, and the same trend is observed in the Figure~\ref{fig3} and  \ref{fig4}. On the other hand, the energy resolution becomes worse as the gain decreases. After 167~hours from the stating point the HV is increased to -~4700~V and all the observables are recorded, some automatically such as spectra, t, p, RH and some manually such as bias current, count rate, gas flow rate.

In this work, the T/p effect on gain and energy resolution is studied in detail. The total data is divided for different time zones at different voltage settings. The gain and energy resolution is plotted as a function of the ratio \( T/p \) (temperature over pressure) across different operational time windows, in Figure~\ref{fig5} and Figure~\ref{fig6} respectively. Exponential relations \( G = A \cdot \exp(B \cdot T/p) \) and \( E_{\text{res}} = A' \cdot \exp(B' \cdot T/p) \) are used respectively to fit the gain and energy resolution as a function of \( T/p \). The fit parameters for both the gain and energy resolution for different operational time windows are summarised in Table~\ref{tab:gain_resolution_fits}. Each row corresponds to a distinct time interval during the detector operation, illustrating how the gain and resolution respond to ambient parameter variations over more than 2200 hour span. Using the \( A \),  \( B \) and \( A' \),  \( B' \) of different zones the measured gain and energy resolution values are normalised to eliminate the variation of those observables due to the variation of T/p.   %The table highlights the consistent increase in gain and the corresponding improvement in energy resolution with increasing \( T/p \), reinforcing the critical role of ambient parameter control in precision measurements and long-term detector stability.

%%%%%%%%%%%%%%%%%%%%%%%%%%%%%%%%%%%%%%%%%%%%%%%%%%%
\begin{figure}[htb!]
	\begin{center} 
		\includegraphics[scale=0.630]{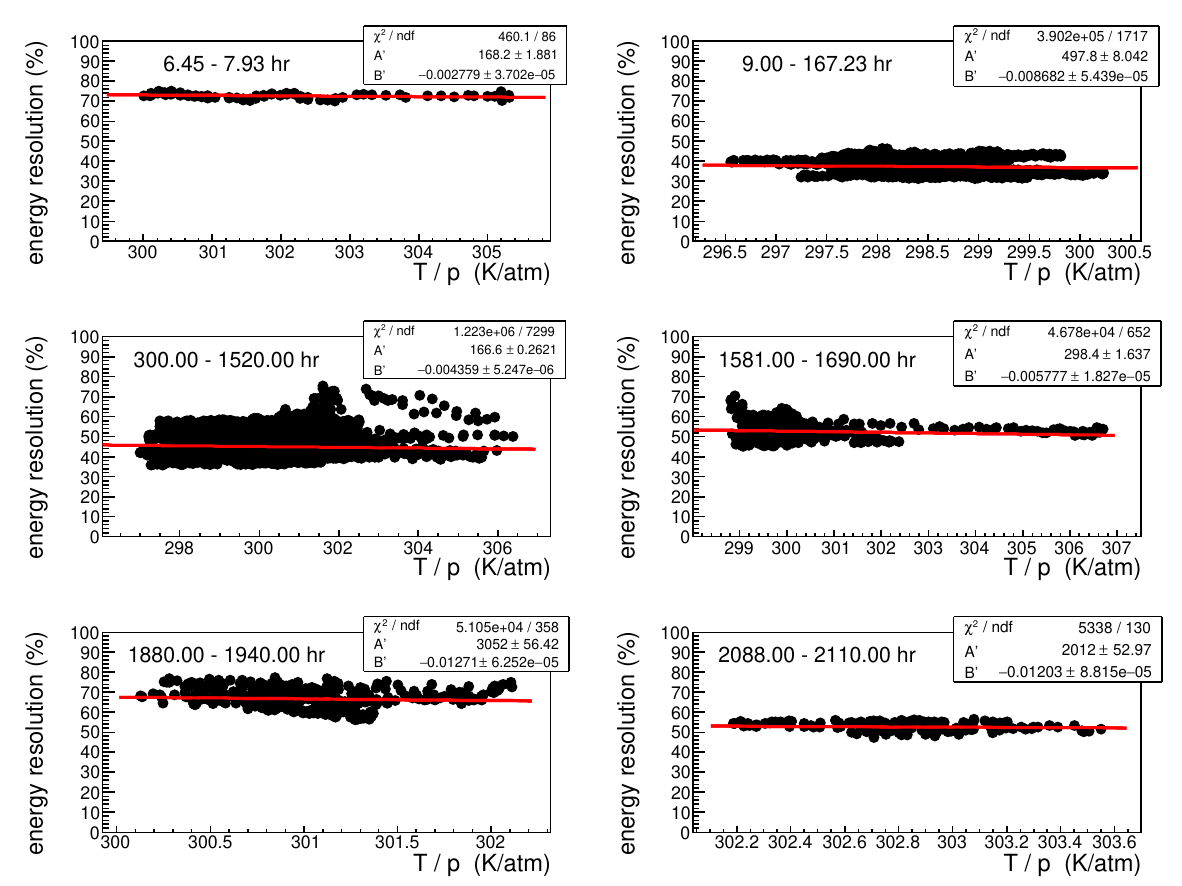}
		\caption{(Colour online) Energy resolution and T/p correlation plot for different time zones. }\label{fig6}
	\end{center}
\end{figure}
%%%%%%%%%%%%%%%%%%%%%%%%%%%%%%%%%%%%%%%%%%%%%%%%%%%

%%%%%%%%%%%%%%%%%%%%%%%%%%%%%%%%%%%%%%%%%%%%%%%%%%%
\begin{table}[h!]
\centering
\begin{tabular}{|c|c|c|c|c|}
\hline
\textbf{Time Interval (hours)} & \( A \) & \( B \) & \( A' \) & \( B' \) \\
\hline
6.45 -- 7.93 & \(63.7 \pm 1.15\) & \(0.0137 \pm 6.002e^{-5}\) & \(168.2 \pm 1.88\) & \(-0.0028 \pm 3.702e^{-5}\) \\
9.00 -- 167.23 & \(9.6 \pm 0.001\) & \(0.0229 \pm 3.15e^{-7}\) & \(497.8 \pm 8.04\) & \(-0.0087 \pm 5.439e^{-5}\) \\
300.00 -- 1520.00 & \(202.4 \pm 0.02\) & \(0.0116 \pm 3.855e^{-7}\) & \(166.6 \pm 0.26\) & \(-0.0044 \pm 5.247e^{-6}\) \\
1581.00 -- 1690.00 & \(127.0 \pm 0.28\) & \(0.0126 \pm 7.428e^{-6}\) & \(298.4 \pm 1.64\) & \(-0.0058 \pm 1.827e^{-5}\) \\
1880.00 -- 1940.00 & \(303.7 \pm 0.48\) & \(0.0094 \pm 5.178e^{-6}\) & \(3052 \pm 56.42\) & \(-0.0127 \pm 6.252e^{-5}\) \\
2088.00 -- 2110.00 & \(225.8 \pm 0.53\) & \(0.0109 \pm 7.508e^{-6}\) & \(2012 \pm 52.97\) & \(-0.0120 \pm 8.815e^{-5}\) \\
\hline
\end{tabular}
\caption{Fit parameters \( A, B \) for gain and \( A', B' \) for energy resolution across different time intervals.}
\label{tab:gain_resolution_fits}
\end{table}
%%%%%%%%%%%%%%%%%%%%%%%%%%%%%%%%%%%%%%%%%%%%%%%%%%%

%%%%%%%%%%%%%%%%%%%%%%%%%%%%%%%%%%%%%%%%%%%%%%%%%%%
\begin{figure}[htb!]
	\begin{center} 
		\includegraphics[scale=0.6]{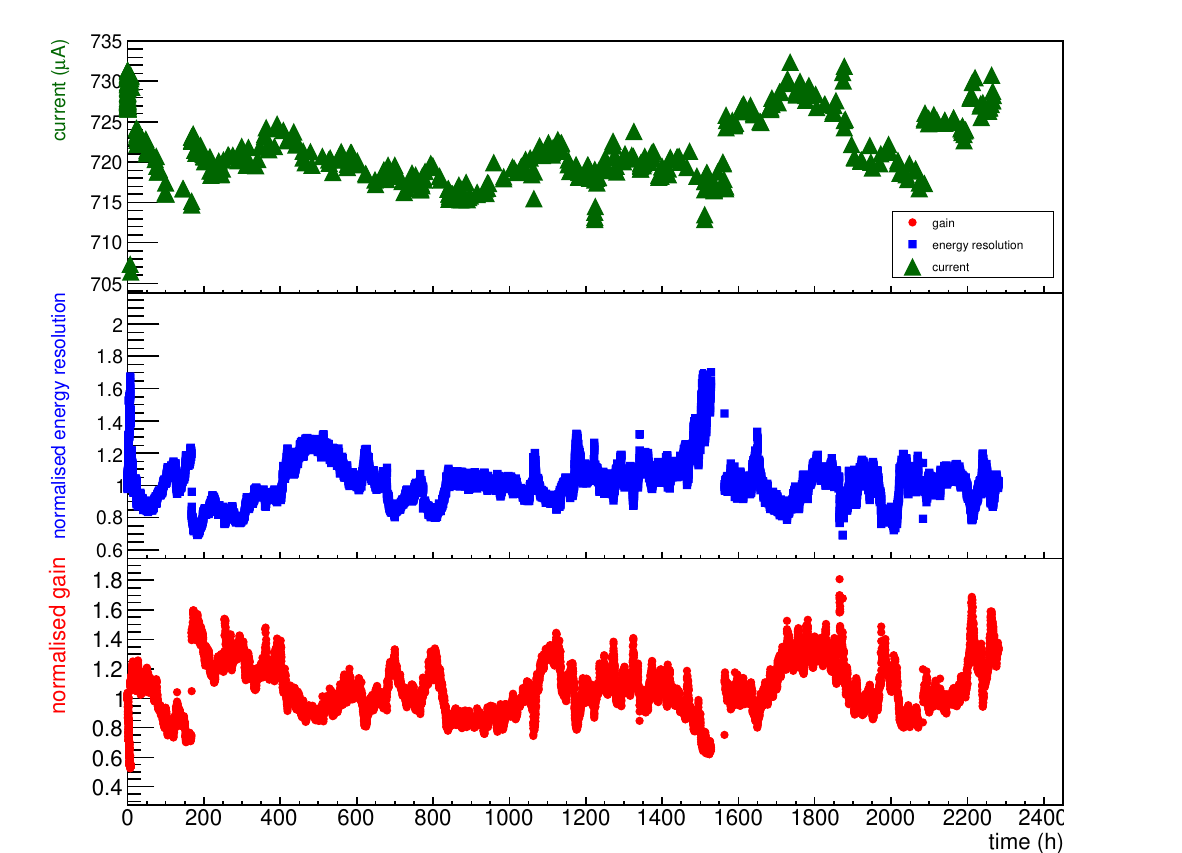}
		\caption{(Colour online) Normalised gain, normalised energy resolution along with the variation of divider current as a function of time.}\label{fig7}
	\end{center}
\end{figure}
%%%%%%%%%%%%%%%%%%%%%%%%%%%%%%%%%%%%%%%%%%%%%%%%%%%

For each data point the measured gain and energy resolution are normalised by the theoretical exponential relations \( G = A \cdot \exp(B \cdot T/p) \) and \( E_{\text{res}} = A' \cdot \exp(B' \cdot T/p) \) respectively, where the fit parameters \( A \),  \( B \) and \( A' \),  \( B' \) are used from Table~\ref{tab:gain_resolution_fits} for different time zones and for different voltage settings. This normalisation eliminates the fluctuation of gain and energy resolution for T/p effect. Figure~\ref{fig7} shows the variation of normalised gain and  normalised energy resolution, along with the bias current for a continuous operation of a period over $\sim$~2200 hours. Finally, the time is converted to accumulated charge per unit irradiated area ($dq$) of the GEM prototype by 

\begin{equation}
\centering
\frac{dq}{dA} = \frac{r \times n \times e \times G \times dt}{dA}\label{chperarea}
\end{equation} 

where, $r$ is the measured rate in Hz incident on a particular area of the detector, $dt$ is the time in second, $n$ is the number of primary electrons for a single X-ray photon, $e$ is the electronic charge, $G$ is the gain and $dA$ is the area of the irradiated part of the chamber. The normalised gain and normalised energy resolution are plotted as a function of charge accumulated per unit area, which is directly proportional to time in Figure ~\ref{fig9}. In the total period of over 2200~hours the charge accumulated per unit area is found to be $\sim$~8.22~mC~mm$^{-2}$ with a continuous X-ray irradiation of rate 220~kHz.

%%%%%%%%%%%%%%%%%%%%%%%%%%%%%%%%%%%%%%%%%%%%%%%%%%%
\begin{figure}[htb!]
	\begin{center} 
		\includegraphics[scale=0.6]{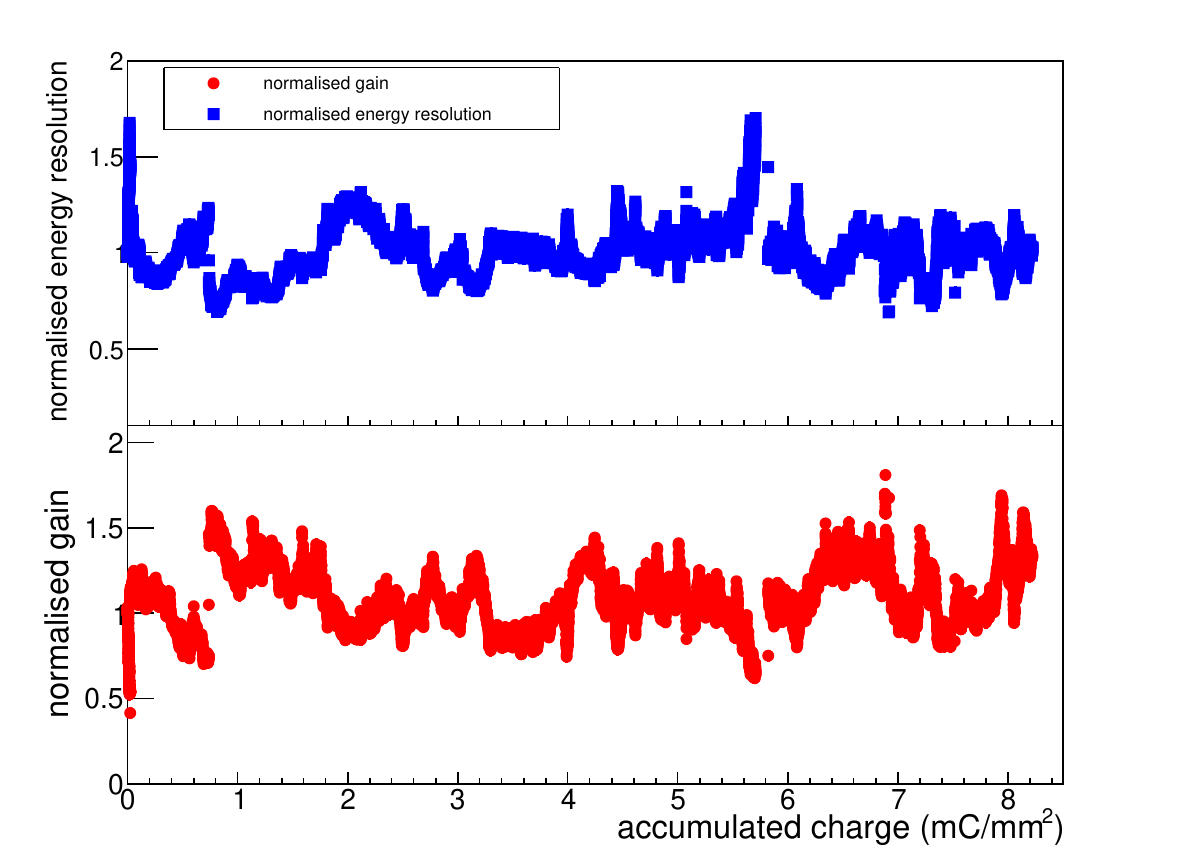}
		\caption{(Colour online) Normalised gain and normalised energy resolution as a function of accumulated charge per unit area.}\label{fig9}
	\end{center}
\end{figure}
%%%%%%%%%%%%%%%%%%%%%%%%%%%%%%%%%%%%%%%%%%%%%%%%%%%

The normalised gain varies from 0.5 to 1.6, while the normalised energy resolution varies between 0.8 and 1.6 primarily due to variation of the bias current. It is observed that the gain initially decreases because of conditioning effect after gain reaches maximum some time later decreases rapidly because of bias voltage varies, Additionally, the Figure~\ref{fig4} shows the energy resolution and gain are anti-correlated. 

%%%%%%%%%%%%%%%%%%%%%%%%%%%%%%%%%%%%%%%%%%%%%%%%%%%
\begin{figure}[htb!]
	\begin{center} 
		\includegraphics[scale=0.6]{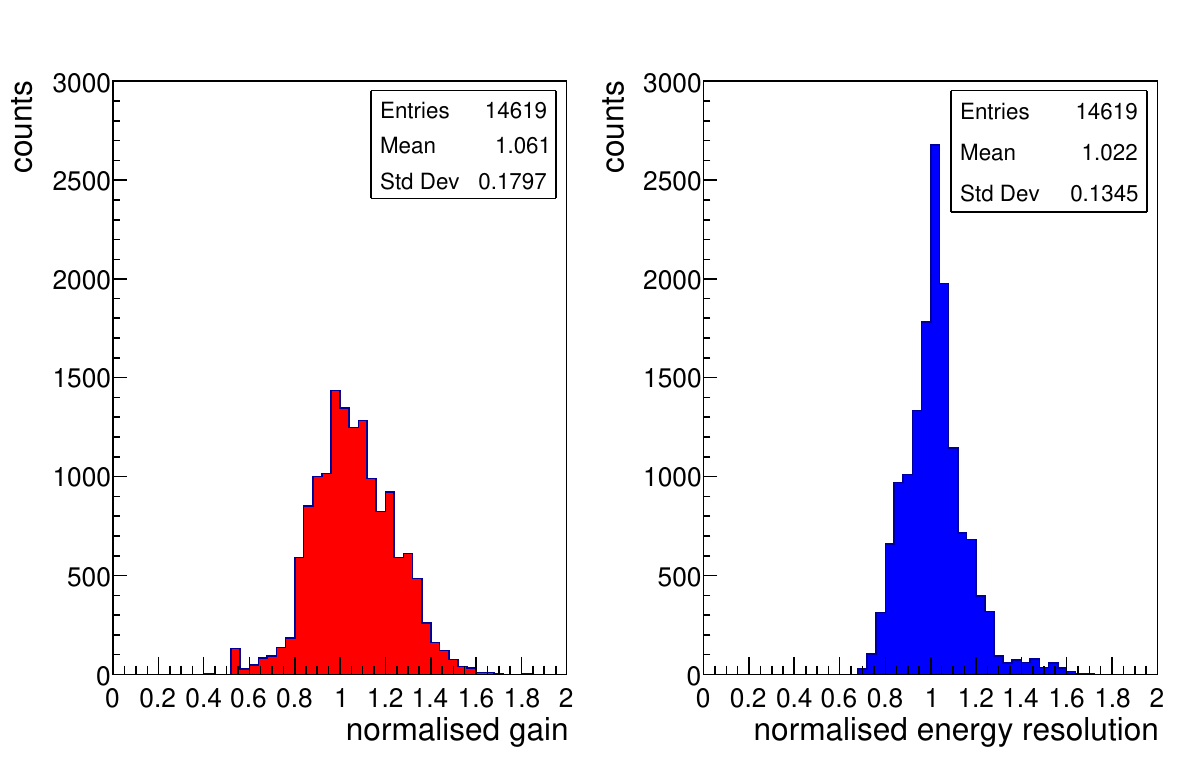}
		\caption{(Colour online) Distribution of normalised gain and normalised energy resolution.}\label{fig8}
	\end{center}
\end{figure}
%%%%%%%%%%%%%%%%%%%%%%%%%%%%%%%%%%%%%%%%%%%%%%%%%%%

%%%%%%%%%%%%%%%%%%%%%%%%%%%%%%%%%%%%%%%%%%%%%%%%%%%
\begin{figure}[htb!]
	\begin{center} 
		\includegraphics[scale=0.43]{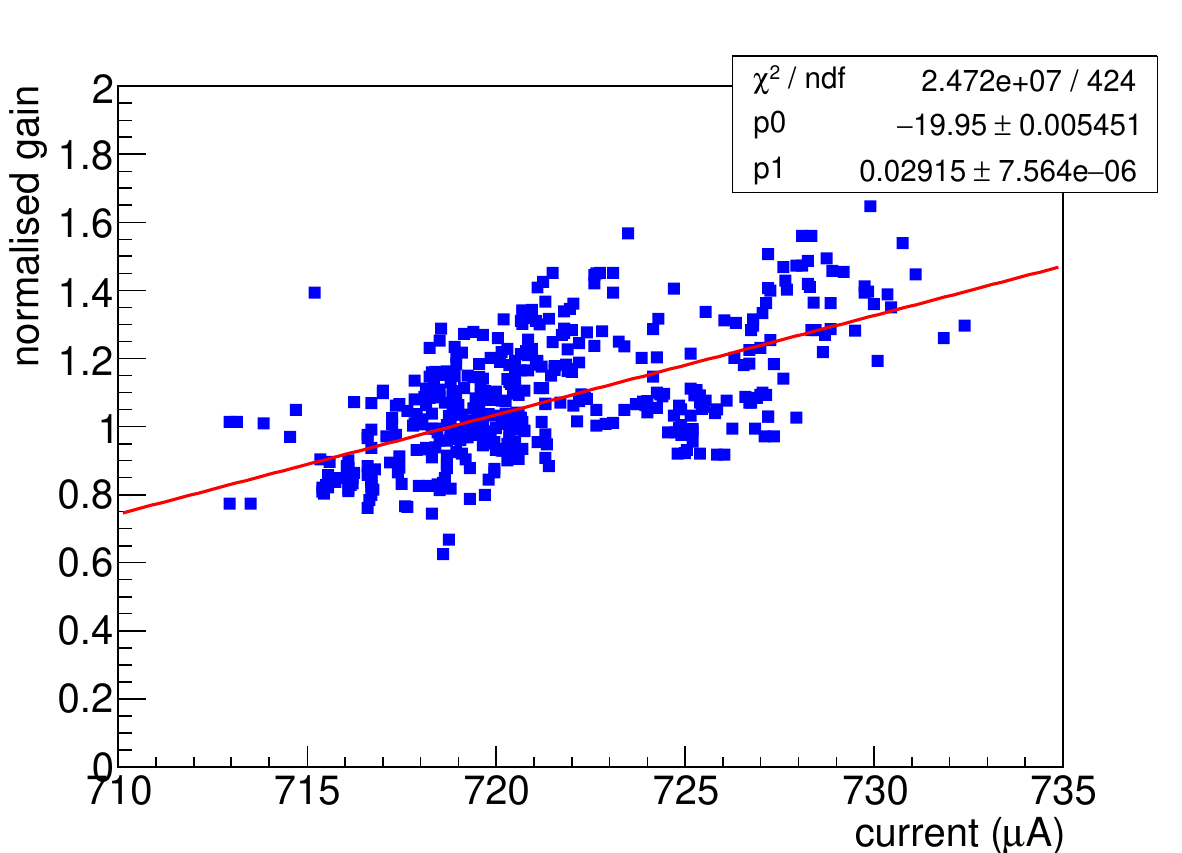}
		\includegraphics[scale=0.43]{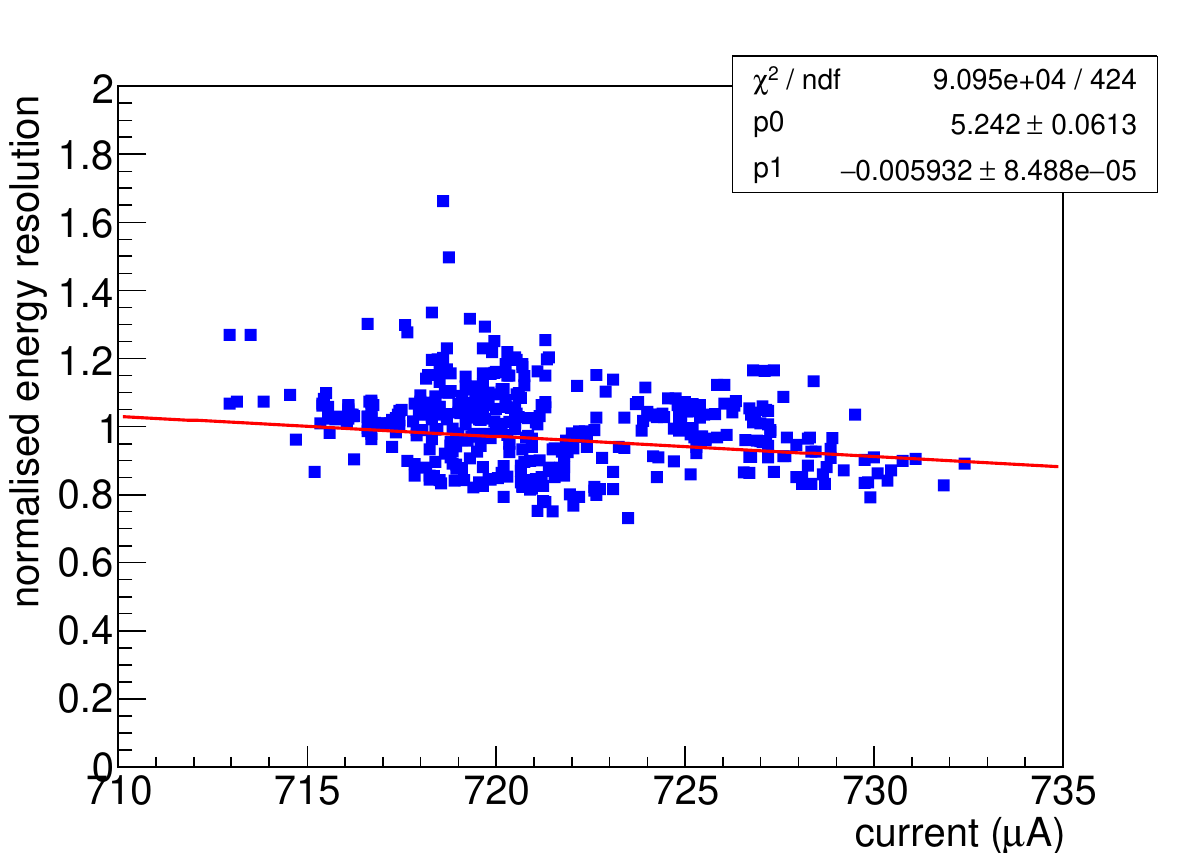}
		\caption{(Colour online) Normalised gain (Left) and normalised energy resolution (Right) as a function of bias current.}\label{fig10}
	\end{center}
\end{figure}
%%%%%%%%%%%%%%%%%%%%%%%%%%%%%%%%%%%%%%%%%%%%%%%%%%%

Figure~\ref{fig8} shows the distributions of normalised gain and normalised energy resolution where the mean values are found to be 1.06~$\pm$~0.18 and 1.02~$\pm$~0.13 respectively, after a charge accumulated per unit area of 8.22~mC~mm$^{-2}$. However, even after continuous irradiation of X-ray of rate 220~kHz for a period over 2200~hours without any single interruption, no continuous decrease of gain or ageing is observed.

It has been observed earlier that, there was correlations between the T/p normalised gain and normalised energy resolution with the bias current  \cite{mandal_2024}. In this continuous long-term study also sometimes the decrease in the bias current was observed (Figure~\ref{fig7}). To check it more clearly, the normalised gain and normalised energy resolution are plotted as a function of the bias current and shown in Figure~\ref{fig10}. The normalised gain and normalised energy resolution vs. bias current plot is fitted by linear function. From Figure~\ref{fig10} the set of fit parameters $p0$, $p1$ and $p0'$, $p1'$ are extracted for normalised gain and normalised energy resolution. The T/p normalised gain and normalised energy resolution are then corrected for the bias current variation using the formulae 

\begin{equation}
\centering
bias~current~corrected~norm.~gain = \frac{normalised~gain}{p0~+~p1~\cdot~bias~current}\label{bcng}
\end{equation} 
and
\begin{equation}
\centering
bias~current~corrected~norm.~energy~resolution = \frac{normalised~energy~resolution}{p0'~+~p1'~\cdot~bias~current}\label{bcner}
\end{equation} 

and once again plotted as a function of charge accumulated per unit area as shown in Figure~\ref{fig11}. The distributions for the bias current corrected normalised gain and bias current corrected energy resolution show  mean values of 1.00~$\pm$~0.14 and 1.05~$\pm$~0.11 respectively. Although it is observed from Figure~\ref{fig10} that the normalised gain increases with current and normalised energy resolution value decreases with current, the bias current corrected normalised gain and normalised energy resolution don't change significantly from the uncorrected normalised gain and normalised energy resolution.

%%%%%%%%%%%%%%%%%%%%%%%%%%%%%%%%%%%%%%%%%%%%%%%%%%%
\begin{figure}[htb!]
	\begin{center} 
		\includegraphics[scale=0.6]{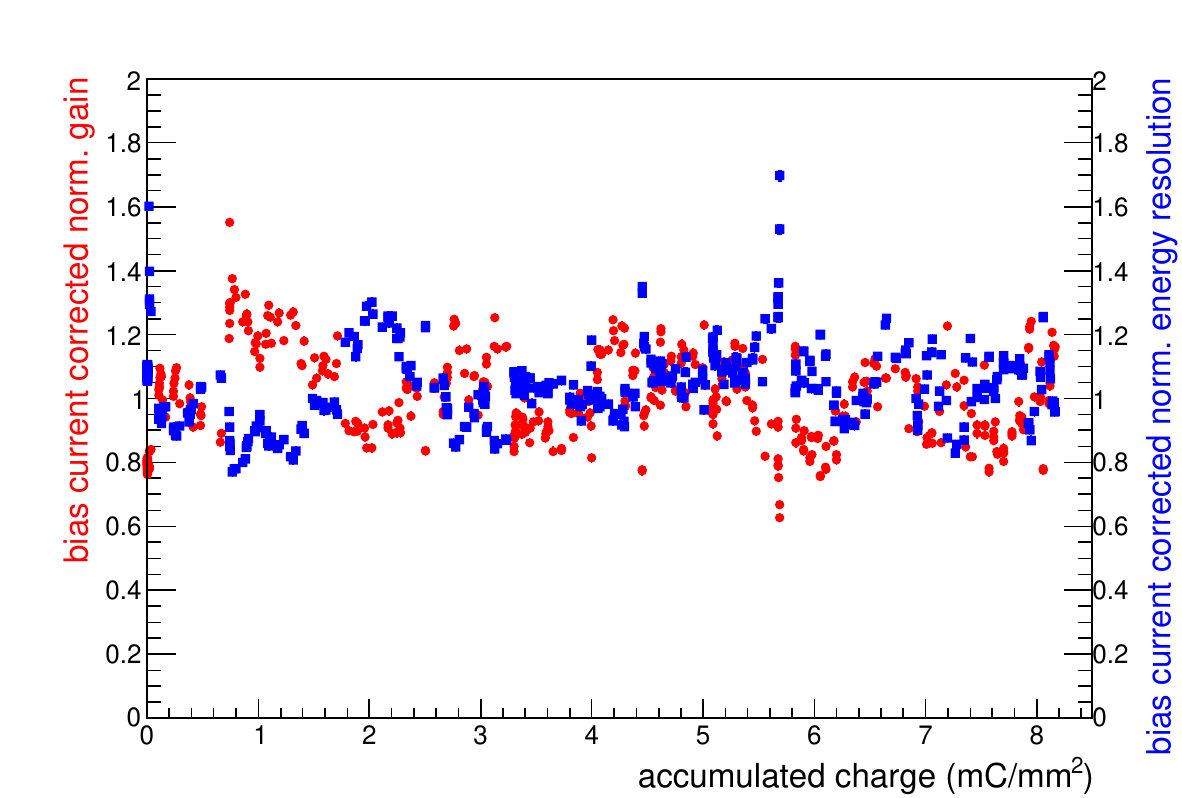}
		\caption{(Colour online) Bias current corrected normalised gain and normalised energy resolution as a function of accumulated charge per unit area.}\label{fig11}
	\end{center}
\end{figure}
%%%%%%%%%%%%%%%%%%%%%%%%%%%%%%%%%%%%%%%%%%%%%%%%%%%
%%%%%%%%%%%%%%%%%%%%%%%%%%%%%%%%%%%%%%%%%%%%%%%%%%%
%\begin{figure}[htb!]
%	\begin{center} 
%		\includegraphics[scale=0.6]{bias_corrected_hist.pdf}
%		\caption{(Colour online) Distribution of bias current corrected normalised gain and normalised energy resolution}\label{fig12}
%	\end{center}
%\end{figure}
%%%%%%%%%%%%%%%%%%%%%%%%%%%%%%%%%%%%%%%%%%%%%%%%%%%

%%%%%%%%%%%%%%%%%%%%%%%%%%%%%%%%%%%%%%%%%%%%%%%%%%%

%%%%%%%%%%%%%%%%%%%%%%%%%%%%%%%%%%%%%%%%%%%%%%%%%%%
\begin{figure}[htb!]
	\begin{center} 
		\includegraphics[scale=0.6]{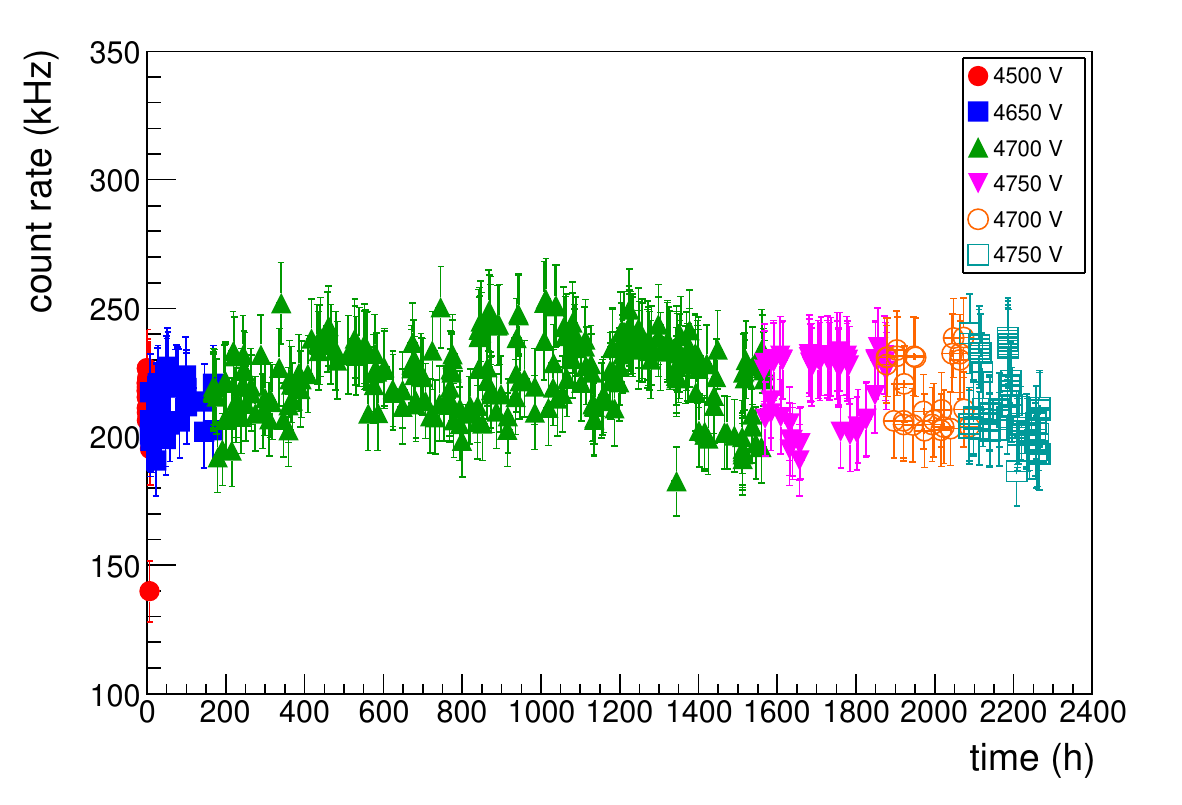}
		\caption{(Colour online) Count rate as a function of time.}\label{fig13}
	\end{center}
\end{figure}
%%%%%%%%%%%%%%%%%%%%%%%%%%%%%%%%%%%%%%%%%%%%%%%%%%%

%%%%%%%%%%%%%%%%%%%%%%%%%%%%%%%%%%%%%%%%%%%%%%%%%%%
\begin{figure}[htb!]
	\begin{center} 
		\includegraphics[scale=0.6]{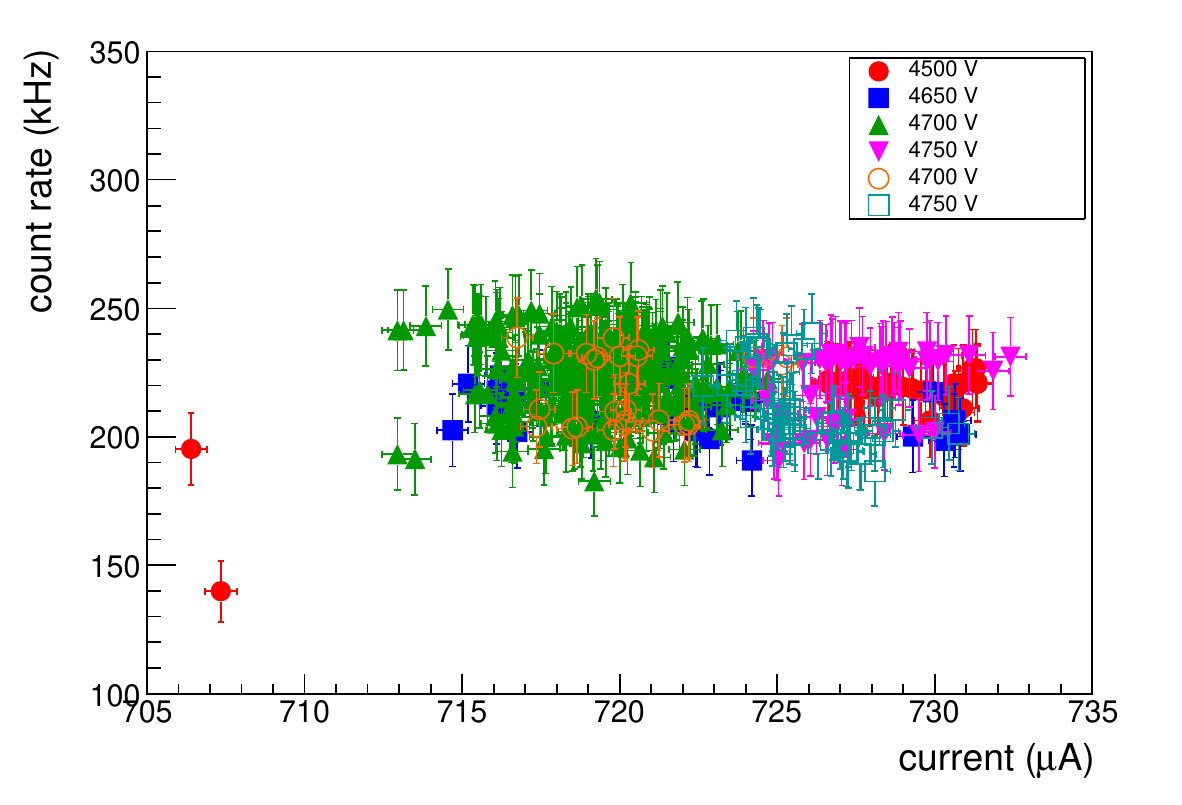}
		\caption{(Colour online) Count rate as a function of bias current.}\label{fig13a}
	\end{center}
\end{figure}
%%%%%%%%%%%%%%%%%%%%%%%%%%%%%%%%%%%%%%%%%%%%%%%%%%%
%%%%%%%%%%%%%%%%%%%%%%%%%%%%%%%%%%%%%%%%%%%%%%%%%%%
\begin{figure}[htb!]
	\begin{center} 
		\includegraphics[scale=0.6]{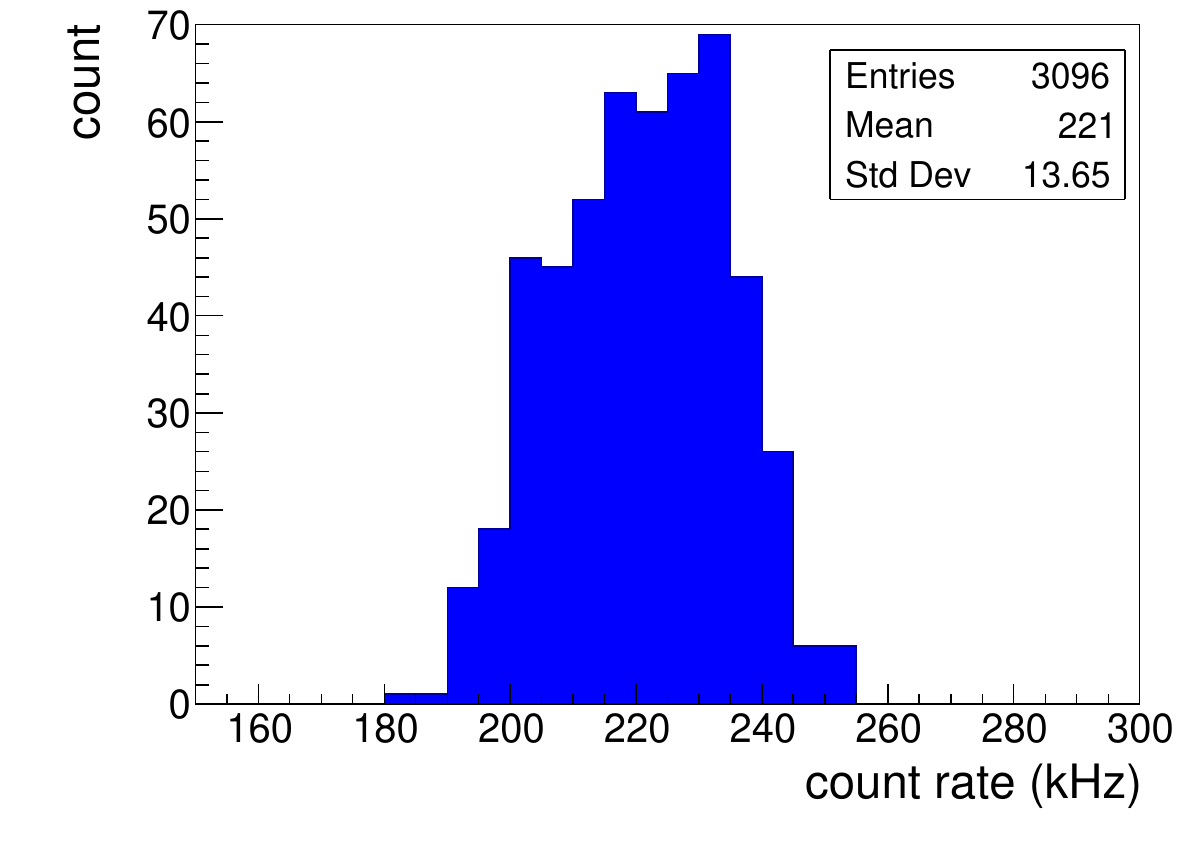}
		\caption{(Colour online) Distribution of the count rate.}\label{fig15}
	\end{center}
\end{figure}
%%%%%%%%%%%%%%%%%%%%%%%%%%%%%%%%%%%%%%%%%%%%%%%%%%%
%%%%%%%%%%%%%%%%%%%%%%%%%%%%%%%%%%%%%%%%%%%%%%%%%%%
\begin{figure}[htb!]
	\begin{center} 
		\includegraphics[scale=0.6]{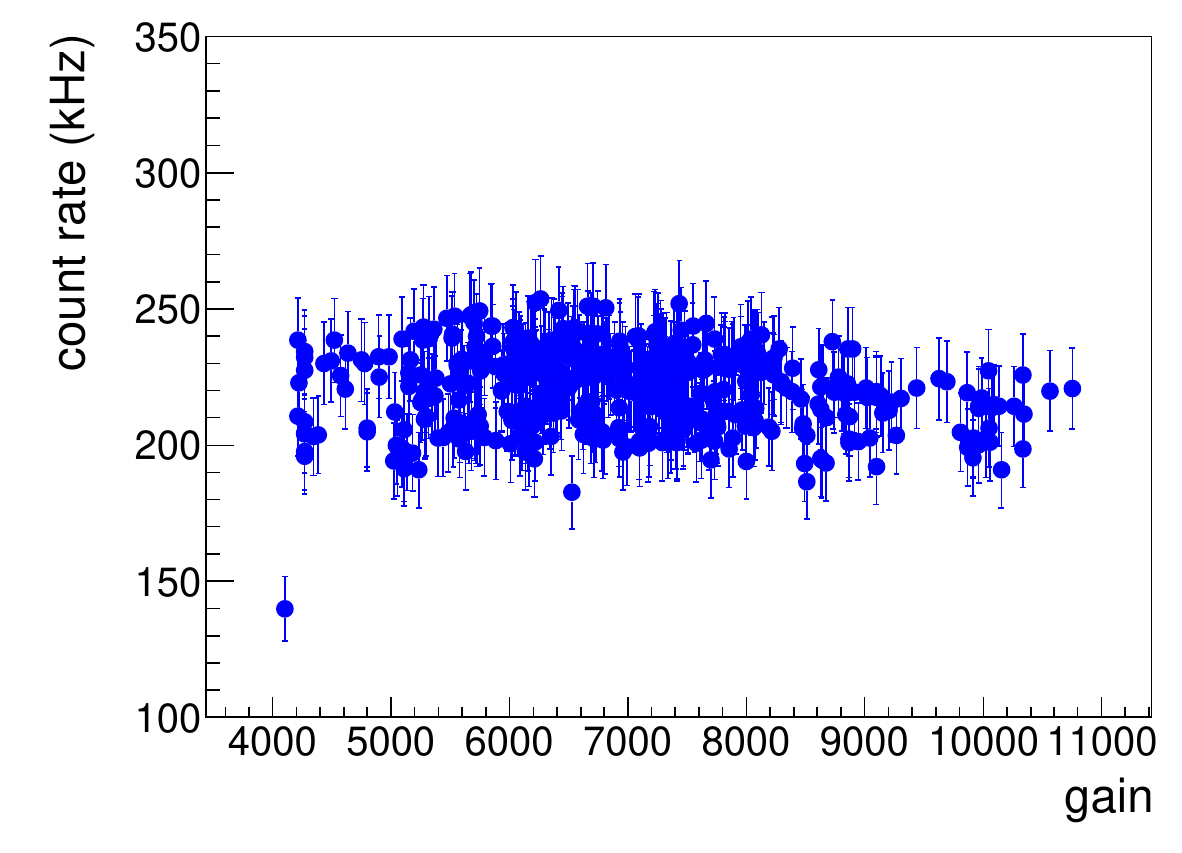}
		\caption{(Colour online) Count rate as a function of gain.}\label{fig14}
	\end{center}
\end{figure}
%%%%%%%%%%%%%%%%%%%%%%%%%%%%%%%%%%%%%%%%%%%%%%%%%%%

Stability in the count rate with a radioactive source can be treated as stability in the efficiency of the detector \cite{mandal_stability}. In this long-term study, in addition to the gain and energy resolution, the variation of the efficiency (count rate with source) with time and current is also studied. In Fig.~\ref{fig13}, the count rate is plotted as a function of time. In this plot different applied voltage settings are shown with different colour and marker. It is observed that during the initial phase of operation, the efficiency increases with time along with the divider current. As the time progresses, although the divider current decrease gradually, the efficiency remains nearly stable. It is important to mention here that the applied voltage is adjusted during the operation to compensate for the drop in current, ensuring a stable $\Delta V$ across the GEM foils. This adjustment contributes to the observed stability of the count rate even during prolonged periods of current decrease.

To investigate the direct correlation between count rate and current, the count rate is plotted as a function of divider current in Fig.\ref{fig13a} for different voltage settings. It is observed that the count rate increases slowly with increasing current, indicating gradual improvement in gain and detection efficiency. However, beyond a certain current threshold, typically above $715\text{--}725\mu\text{A}$, the efficiency remains almost stable. This behavior is more pronounced at higher voltage settings, such as $4700\text{--}4750$~V, where the count rate is found to be between $200\text{--}240,\text{kHz}$. The stable efficiency recorded is approximately $220~\text{kHz}$ at a divider current of $\sim~720\mu\text{A}$ for applied voltage of $4750$~V, which can be interpreted as the detector reaching its maximum operational efficiency. The distribution of count rate Fig.\ref{fig15} shows that the average count rate is $221~\text{kHz}$ with a standard deviation of 13.6~kHz.

The count rate as a function of the measured gain is also plotted in Fig.\ref{fig14}. No significant correlation between count rate and gain is observed over the studied gain range of 4000 to 11000. This indicates that the detector efficiency does not change significantly with the change in the gain.

\section{Summary and Discussion}
This study explored the performance features of a triple GEM detector emphasising its gain, energy resolution and count rate as function of time or equivalently charge per unit area. The effect of the mentioned observables on the ambient parameters are also studied. The measurement is carried out uninterruptedly over 90 days. The high voltage was on and 50~mm$^2$ of the chamber was exposed with constant radiation of X-ray at a rate of 220~kHz. In this continuous study of over 90 days or after a charge accumulated per unit area of 8.22~mC~mm$^{-2}$, the T/p normalised gain and normalised energy resolution remain constant at a mean values of 1.06~$\pm$~0.18 and 1.02~$\pm$~0.13 respectively.\par

During the measurement it is observed that sometimes the divider current decreases slowly over time, to keep the $\Delta V$ constant the high voltage is varied manually time to time. Even after renormalisation of these values for the variation of the bias current the corrected normalised gain and corrected energy resolution show mean values of 1.00~$\pm$~0.14 and 1.05~$\pm$~0.11 respectively. No continuous degradation is observed even after this continuous uninterrupted radiation.\par

The count rate of the chamber, measured with the radioactive source as an indicator of operational efficiency and it is carefully monitored throughout the period of operation. It is observed that in the initial phase of operation, particularly during the first few hours, the count rate increases along with the divider current as the applied voltage is gradually raised. However, after reaching the target operating conditions, the divider current shows a slow declining trend over time, whereas the count rate remains nearly stable. This initial behaviour can be attributed to a combination of the detector conditioning phase and charge stabilisation effects within the dielectric layers of the GEM foils.

The effect of the count rate on the gain is also studied. Other than the conditioning phase, the results don't show any correlation between the gain and  the count rate. The count rate is more or less constant over the entire period of operation. Such trend reaffirms the sensitivity in the amplification characteristics of GEM based detectors and shows that the efficiency remains nearly the same over a large gain value. These understandings are important in determining the operational limits in high accuracy and stability applications.\par

From this study, it can be concluded that while gain and energy resolution display gradual fluctuations with current and time, the count rate representing the efficiency remains stable after the initial conditioning phase other than a fluctuation from the mean value. This operational stability, specially in terms of efficiency, is a critical feature for applications of such a tracking detectors, where long-term performance without frequent recalibration is necessary. The key take home message and the unique feature of this study is that, there is no ageing observed in GEM detector after a long-term continuous study over 90 days when the bias was on and the detector was continuously kept under strong source of irradiation.

%%\label{}
%%\lipsum[4]

%\section{Summary and conclusions}
%%\label{}
%\lipsum[1-4]

%\section*{Acknowledgements}
\section{Acknowledgements}

The authors would like to thank the RD51 collaboration for the support in building and initial testing of the chamber in the RD51 laboratory at CERN.The authors would like to thank Dr. Arindam Sen of JLab, USA and Dr. Sayak Chatterjee of UMass for valuable discussions in this work. also would like to thanks Ms. Manika Agarwal to help data taking. We would also like to thank Mr. Subrata Das for helping in building of the collimators used in this study. This work is partially supported by the research grant SR/MF/PS-01/2014-BI from DST, Govt. of India, and the research grant of the CBM-MuCh project from BI-IFCC, DST, Govt. of India. S. Mandal acknowledges his UGC-NET fellowship for the support. S. Biswas acknowledges the support of the DST-SERB Ramanujan Fellowship (D.O.No. SR/S2/RJN-02/2012).

%\appendix

%\section{Appendix title 1}

%\section{Appendix title 2}

%\bibliographystyle{elsarticle-harv} 
%\bibliography{example}
%\bibliography{}

\begin{thebibliography}{9}
	\bibitem{Sauli} F. Sauli, Nucl. Instrum. Methods Phys. Res. A 386 (1997) 531.
	\bibitem{Buzulutskov} A.F. Buzulutskov, Instrum Exp Tech 50, (2007) 287.
	\bibitem{Ketzer} B. Ketzer \textit{et al.,} , Nucl. Instrum. Methods Phys. Res. A 535 (2004) 314.
	\bibitem{Galatyuk} T. Galatyuk, Nucl. Phys. A 982 (2019) 163.
	\bibitem{cbm} https://www.cbm.gsi.de/
	\bibitem{fair} http://www.fair-center.eu/
	\bibitem{rama} Rama Prasad Adak  \textit{et al.,} Nucl. Instrum. Methods Phys. Res. A 846 (2017) 29.
	\bibitem{sharma2023} A. K. Sharma \textit{et al.,} Proc. DAE Symp. Nucl. Phys. 67 (2023) 1011.
	\bibitem{biswas2015} S. Biswas \textit{et al.,} Nucl. Instrum. Methods Phys. Res. A 800 (2015) 93.
	\bibitem{oliveira2013} Oliveira \textit{et al.,} United States Patent, US 8,597,490 B2 (2013).
	\bibitem{duarte2009} S. Duarte Pinto \textit{et al.,} JINST 4 (2009) P12009.
	\bibitem{karadzhinova2015} A. Karadzhinova \textit{et al.,} JINST 10 (2015) P12014.
	\bibitem{das2016} S. Das, Nucl. Instrum. Methods Phys. Res. A 824 (2016) 518.
	\bibitem{shah2019} A. Shah \textit{et al.,} Nucl. Instrum. Methods Phys. Res. A 936 (2019) 459.		
	\bibitem{bachmann1999} S. Bachmann \textit{et al.,} Nucl. Instrum. Methods Phys. Res. A 438 (1999) 376.
	\bibitem{s_chatterjee_charging_up_2} S. Chatterjee \textit{et al.,} Nucl. Instrum. Methods Phys. Res. A 1014 (2021) 165749.
	\bibitem{mandal_stability} S. Mandal \textit{et al.,} Pramana - Journal of Physics (In Press), arXiv:2505.03357 (2025).
	\bibitem{preamplifier} CDT CASCADE Detector Technologies GmbH, Germany, www.n-cdt.com 
	\bibitem{mandal_2024} S. Mandal \textit{et al.,} Nucl. Instrum. Methods Phys. Res. A 1064 (2024) 169389.	
	\bibitem{Roy_thesis} Shreya Roy, Characterization Of Gaseous And Scintillator Detectors For High Energy Physics And Cosmic Ray Experiments, Ph.D. Thesis, University of Calcutta, 2023.
	\bibitem{Sen_thesis} Arindam Sen, Development Of Resistive Plate Chamber For The CBM Experiment At FAIR And Other Application Of Radiation Detector, Ph.D. Thesis, University of Calcutta, 2023.
	\bibitem{Chatterjee_thesis} Sayak Chatterjee, Performance Studies of Gas Electron Multiplier Detector for the Muon Chamber of High Rate CBM Experiment at FAIR, Ph.D. Thesis, University of Calcutta, 2023.	
	\bibitem{adak} R.P. Adak \textit{et al.,} 2016 JINST 11 T10001. 
	\bibitem{s_roy_gain_calculation} S. Roy \textit{et al.,} Nucl. Instrum. Methods Phys. Res. A 936 (2019) 485.
	\bibitem{uniformity_1} S. Chatterjee \textit{et al.,} Nucl. Instrum. Methods Phys. Res. A 936 (2019) 491.
	\bibitem{chatterjee_2020} S. Chatterjee \textit{et al.,} Journal of Physics: Conference Series 1498 (2020) 012037.
	\bibitem{s_chatterjee_charging_up_1} S. Chatterjee \textit{et al.,} 2020 JINST 15 T09011.	
	\bibitem{chatterjee_2023_rh} S. Chatterjee \textit{et al.,} Nucl. Instrum. Methods Phys. Res. A 1046 (2023) 167747.
	\bibitem{chatterjee_2023_charge} S. Chatterjee \textit{et al.,} Nucl. Instrum. Methods Phys. Res. A 1049 (2023) 168110.
	\bibitem{sahu} S Sahu \textit{et al.,} JINST 12, C05006 (2017). 
	\bibitem{mandal_datalogger} S. Mandal \textit{et al.,} Proc. DAE Symp. Nucl. Phys. 69 (2025) 1299-1300.	
		
	
	
	%\bibitem[]{}
	%\bibitem{paper1} FSauli, Nucl. Instrum. Methods Phys. Res. A 386 (1997) 531.
	
	%\bibitem{F. Sauli, Nucl. Instrum. Methods Phys. Res. A 386 (1997) 531}
	%% For example:
	
	%\bibitem[Aladro et al.(2015)]{Aladro15} Aladro, R., Martín, S., Riquelme, D., et al. 2015, \aas, 579, A101
	
	
\end{thebibliography}

%\section{References}

\end{document}